\documentclass[aps,prb,twocolumn,showpacs,superscriptaddress,floatfix]{revtex4-2}
\usepackage{graphicx}
\usepackage{amsmath}
\usepackage{amssymb}
\usepackage{dcolumn}
\usepackage{bm}
\usepackage{siunitx}
\usepackage{hyperref}
\usepackage{xcolor}
\usepackage{float}
\hypersetup{colorlinks=true,citecolor={blue},linkcolor={blue},urlcolor={blue}}
\usepackage[caption=false]{subfig}
\newcommand{\phantomsubfloat}[1]{
    {
        \captionsetup[subfigure]{labelformat=empty}
        \subfloat[][]{#1}
    }
}
\usepackage{soul}
\setul{1ex}{2ex} 
\definecolor{applegreen}{rgb}{0.55, 0.71, 0.0}

\setstcolor{red}
\usepackage{cancel} 

\begin{document}

\title{Enhanced coupling between ballistic exciton-polariton condensates through tailored pumping}

\author{Y. Wang}
\address{School of Physics and Astronomy, University of Southampton, Southampton SO17 1BJ, United Kingdom}

\author{P. G. Lagoudakis}
\address{Hybrid Photonics Laboratory, Skolkovo Institute of Science and Technology, Territory of Innovation Center Skolkovo, Bolshoy Boulevard 30, Building 1, 121205 Moscow, Russia}
\address{School of Physics and Astronomy, University of Southampton, Southampton SO17 1BJ, United Kingdom}

\author{H. Sigurdsson}
\email{h.sigurdsson@soton.ac.uk}
\affiliation{School of Physics and Astronomy, University of Southampton, Southampton SO17 1BJ, United Kingdom}
\affiliation{Science Institute, University of Iceland, Dunhagi 3, IS-107, Reykjavik, Iceland}

\begin{abstract}
We propose a method to enhance the spatial coupling between ballistic exciton-polariton condensates in a semiconductor microcavity based on available spatial light modulator technologies. Our method, verified by numerically solving a generalized Gross-Pitaevskii model, exploits the strong nonequilibrium nature of exciton-polariton condensation driven by localized nonresonant optical excitation. Tailoring the excitation beam profile from a Gaussian into a polygonal shape results in refracted and focused radial streams of outflowing polaritons from the excited condensate which can be directed towards nearest neighbors. Our method can be used to lower the threshold power needed to achieve polariton condensation and increase spatial coherence in extended systems, paving the way towards creating extremely large-scale quantum fluids of light.
\end{abstract}

\maketitle

\section{Introduction}
Ever since the demonstration of Bose-Einstein condensation of exciton-polaritons (from here on {\it polaritons}) in planar semiconductor microcavities~\cite{kasprzak2006bose} there has been tremendous effort dedicated to scaling up the number of coupled condensates to form extended systems. The notable candidates for large-scale networks and lattices of polariton condensates are etched micropillar arrays~\cite{whittaker2018exciton, Goblot_PRL2019}, metal deposited cavity surface~\cite{Lai_Nature2007}, etch-and-overgrowth techniques~\cite{Su_NatPhys2020, Harder_ACSPho2021}, surface acoustic waves~\cite{Cerda_PRL2010}, and structured nonresonant light source using spatial light modulators~\cite{Tosi_NatComm2012, topfer2021engineering}. On one hand, designing lattices of polariton condensates can offer new insight into the non-Hermitian physics of driven-dissipative quantum fluids obeying Bloch's theorem with strong nonlinearities~\cite{Boulier_AdvQuaTech2020}. On the other hand, the large state space and strong nonlinearities of a coupled polariton condensate network could offer a platform for classical or even quantum computing protocols~\cite{Kavokin_NatRevPhys2022} given the ease of optical write-in and read-out of polaritons (being part photonic)~\cite{RevModPhys.85.299}.

A common challenge in designing extended polariton condensate systems is making the inter-node coupling strong enough to overcome the detrimental effects of disorder and noise which would otherwise reduce the system's coherence. Good coherence in polariton systems can give access to both intricate long-range and long-time condensate dynamics, and also play a role in various optical applications such as biological imaging~\cite{huang1991optical}, information processing~\cite{hotate1994optical, chen2016experimental}, neuromorphic computing~\cite{Shastri_NatPho2021}, and metrology~\cite{giovannetti2004quantum}. 

In this paper, we propose an all-optical method to enhance the spatial coupling and coherence between nonresonantly driven ballistic polariton condensates~\cite{Topfer_CommPhys2020}. To achieve this, we numerically model tailored pump spots with reduced rotational $C_n$ symmetry which generate, refract, and focus high-momentum condensate polariton waves between nearest neighbors. This can be readily achieved in practice using liquid crystal spatial light modulators~\cite{Abmann_PRB2012}. As a case study, we demonstrate our idea by numerically solving the generalized stochastic Gross-Pitaevskii equation for a honeycomb lattice tiled with triangular pump spots ($C_3$) in comparison to more conventional cylindrically symmetric Gaussian spots. Similar structures have been exploited in photonic crystal slabs to generate band gaps~\cite{Takayama_APL2005}, but have not been widely explored in the context of polariton fluids. Our method can be applied to optically driven lattices of polariton condensates which have today reached hundreds of coherently coupled condensates~\cite{topfer2021engineering}. Our results could advance the performance of polariton platforms to explore $XY$ spin materials~\cite{Berloff_NatMat2017, tao2022halide}, topological physics~\cite{pickup2020synthetic, Pieczarka_Optica2021}, vorticity~\cite{Tosi_NatComm2012, Cookson_NatComm2021}, and band structure engineering~\cite{Zhang_Nanoscale2018, alyatkin2020optical}.

\section{Anisotropic pump shapes for polariton condensation}
\label{Anisotropy_in polariton_condensates}
The exciton-polariton condensate is denoted a macroscopic wavefunction (i.e., order parameter) $\Psi(\mathbf{r},t)$ which is governed by a generalized Gross-Pitaevskii equation. The condensate is coupled to a background reservoir density of incoherent excitons $n_X(\mathbf{r},t)$ which are driven by an external nonresonant excitation source $P(\mathbf{r})$ and described by a simplistic rate equation model~\cite{PhysRevLett.99.140402},
\begin{eqnarray}
i\hbar\frac{\partial \Psi}{\partial t}&=&\bigg\{-\frac{\hbar^{2}\nabla^{2}}{2m}
 +\alpha |\Psi|^2 +G\Big[n_X + \frac{\eta }{\Gamma_{X}}P(\mathbf{r})\Big] \nonumber 
\\
&&  +\frac{i \hbar}{2}\left(R n_X-\gamma\right)\bigg\}\Psi,
\label{gross-pitaevskii_equation}
\\
\frac{\partial n_X}{\partial t}&=&-\left(\Gamma_X+R |\Psi|^{2}\right)n_X + P(\mathbf{r}).
\label{rate_equation_of_reservoir}
\end{eqnarray}
Here, $m$ is the effective polariton mass, $G = 2 g |\chi|^2$ and $\alpha = g |\chi|^4$ are, respectively, the repulsive (defocusing) polariton-reservoir and polariton-polariton interaction strengths, $|\chi|^2$ is the excitonic Hopfield coefficient of the polariton, $g$ refers to the exciton-exciton dipole interaction strength, $R$ stands for the rate of stimulated scattering of reservoir excitons into the condensate, $\gamma$ and $\Gamma_X$ are, respectively, the polariton and reservoir decay rates, and $\eta$ determines the ratio of the contribution of the blueshift from both dark and high-momentum excitons which do not scatter into the condensate but are present in the background. We set the parameters similar to our previous works based on slightly negatively detuned cavities with InGaAs quantum wells: $m=0.28\mathrm{\,meV\,ps^{2}\,\mu m^{-2}}$, $|\chi|^{2}=0.4$, $g= 1 \, \mu\mathrm{eV\,\mu m^{2}}$, $\hbar R=2.0g$, $\eta=2$, and $\gamma^{-1} = \Gamma_{X}^{-1} = \SI{5.5}{ps}$.
The nonresonant pump is written $P(\mathbf{r}) = P_0 f(\mathbf{r})$, where $P_0$ is the power density multiplied with a spatial profile satisfying $\text{max}{(f)}=1$.



\section{Shaping the polariton outflow}
We are interested in the steady-state solutions of Eq.~\eqref{gross-pitaevskii_equation} using pump profiles $f(\mathbf{r})$ tailored to guide the polariton waves into desired patterns. This is made possible because a local pumping region (i.e., spot) produces a co-localized complex potential landscape felt by the generated polaritons. In the low-density regime (i.e., close to condensation threshold), this potential is written as
\begin{equation}
V(\mathbf{r})=\frac{P(\mathbf{r})}{\Gamma_{X}}\left[(1+\eta)G+i\hbar\frac{R}{2}\right].
\label{complex_potential}
\end{equation}
From the above equation, one can appreciate two things: (i) The real part is positive because excitons interact repulsively ($G>0$), which means that polaritons are blueshifted at the spot location. (ii) The imaginary part is also positive, which means that above a certain critical power $P_0 = P_\text{th}$, the condensation threshold is reached (stimulation exceeds losses) and coherent polaritons are amplified at the spot location until the reservoir clamps and the condensate stabilizes.

\begin{figure}[t]
\includegraphics[scale=1.0]{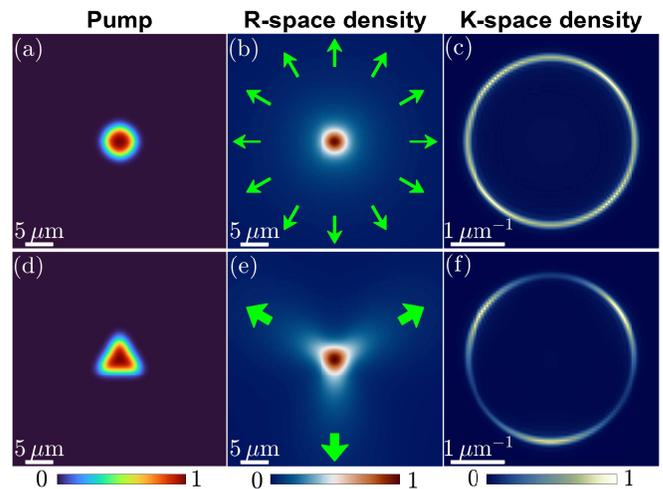}
\phantomsubfloat{\label{fig_isotropy_a}}
\phantomsubfloat{\label{fig_isotropy_b}}
\phantomsubfloat{\label{fig_isotropy_c}}
\phantomsubfloat{\label{fig_isotropy_d}}
\phantomsubfloat{\label{fig_isotropy_e}}
\phantomsubfloat{\label{fig_isotropy_f}}
\vspace{-2\baselineskip}
\caption{(a),(d) Circular and triangular pump configurations and corresponding normalized condensate steady-state solutions $|\Psi|^{2}$ in (b),(e) real space and in (c),(f) momentum space. The green arrows in (b) illustrate the condensate flow emitted in all directions, and in (e), the thicker green arrows show the anisotropic and concentrated condensate flow. Notice how the triangular-shaped condensate in (e) is rotated by $\pi/3$ with respect to the pump in (d).}
\label{fig_isotropy}
\end{figure}

For small pumping spots the resulting steady state is a {\it ballistic condensate}~\cite{Wertz_NatPhys2010, Christmann_PRB235303}, shown in Fig.~\ref{fig_isotropy}. For cylindrically symmetric spots [see Figs.~\ref{fig_isotropy_a}--\ref{fig_isotropy_c}], the real-space condensate density is co-localized with the spot and coherent polariton waves are radially emitted in all directions with high momentum, as evidenced from the sharp density ring in momentum space. When multiple small spots are pumped and displaced from each other, fascinating phenomena such as spontaneous synchronization and multimodal emission can occur due to the complex non-Hermitian coupling between neighboring condensates~\cite{Cristofolini_PRL2013,Topfer_CommPhys2020,alyatkin2020optical, topfer2021engineering}. The condensate coupling is mediated by propagating polariton waves in the plane of the cavity (near field) and should not be confused with out-of-plane coupling like in the far field of laser arrays~\cite{kao2016phase}.

If the pump spot is, however, not cylindrically symmetric, then the generated polariton waves will experience refraction and interference, giving rise to anisotropic streams of condensate polaritons, as shown in Figs.~\ref{fig_isotropy_d}--\ref{fig_isotropy_f} for a triangular spot. Such shaping of the pump spot can be realized using spatial light modulators on the incident excitation, which allows focusing almost arbitrary excitation patterns onto the microcavity plane~\cite{Abmann_PRB2012}. This has enabled the demonstration of all-optical in-plane polariton waveguides~\cite{PhysRevB.91.195308, Cristofolini_PRL2018}, transistor switches~\cite{Gao_PRB2012, Anton_PRB2013}, amplification~\cite{Niemietz_PRB2016}, tailored momentum distribution~\cite{Abmann_PRB2012}, and microlensing~\cite{wang2021reservoir}. Notice how the triangular-shaped condensate in Fig.~\ref{fig_isotropy_e} is rotated by $\pi/3$ with respect to its pump pattern in Fig.~\ref{fig_isotropy_d}. This can be understood from the following consideration. Inside the pump spot, low-momentum polariton waves $(k \sim 0)$ are amplified and subsequently disperse, flowing out of the pump spot and elastically converting their (pump-induced) potential energy into kinetic energy. The waves which hit the edges of the triangle close to normal incidence scatter very little, whereas polaritons hitting the corners of the triangle are at an oblique incidence and scatter more strongly. This leads to enhanced flow of condensate particles along the normals of triangle edges, effectively forming the dual pattern of the pump (i.e., pump edges map to condensate corners). The same holds for higher-order polygonal-shaped pump spots.
This interpretation can be easily verified by solving an initial value problem of a two-dimensional Schr\"{o}dinger equation wherein a Gaussian wave packet centered at $k=0$ and $\mathbf{r}=0$ in momentum and real space is propagated in time. We will show in the following that shaping multiple spots into triangles, as opposed to the conventional Gaussian-shaped spots, focuses and enhances the interaction between adjacent condensates, resulting in lowered threshold and increased coherence in the extended polariton system.  

Since different pumping profiles usually have different condensation threshold power density, the triangular spot needs to be calibrated against the circular (Gaussian) spot so that they share the same threshold power density, $P_{\text{th},T} = P_{\text{th},C}$. Under this condition, polaritons experience the same blueshift at their respective spots and thus populate the momentum components of similar magnitude in Fourier space [compare Fig.~\ref{fig_isotropy_c} with~\ref{fig_isotropy_f}], enabling a fairer comparison. This calibration is achieved by fixing the parameters of the model~\eqref{gross-pitaevskii_equation} and adjusting the side length of the triangle until $P_{\text{th},T} = P_{\text{th},C}$ for a given full width at half maximum of the circular spot.

\begin{figure}[t]
\includegraphics[scale=1.0]{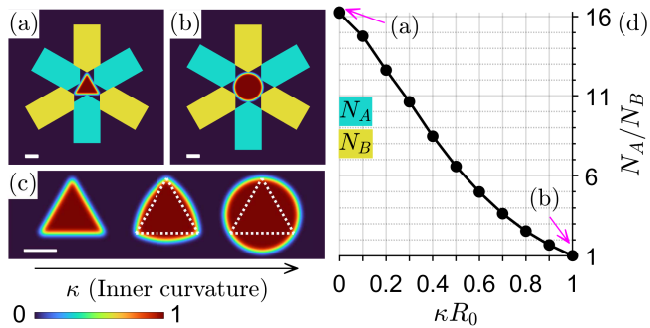}
\phantomsubfloat{\label{fig_particles_a}}
\phantomsubfloat{\label{fig_particles_b}}
\phantomsubfloat{\label{fig_particles_c}}
\phantomsubfloat{\label{fig_particles_d}}
\phantomsubfloat{\label{fig_particles_e}}
\phantomsubfloat{\label{fig_particles_f}}
\vspace{-2\baselineskip}
\caption{(a),(b) Triangular and circular pump profiles (red color) with overlaid schematic blue and yellow integration areas to determine the amount of condensate anisotropy in the system. (c) Example of pump profiles with zero, intermediate, and maximal inner curvature denoted $\kappa=1/R$. The white dotted line indicates the circumscribed triangle. (d) The ratio of condensate particles (integrated density) between the blue and yellow areas $N_A/N_B$ for varying curvature for a fixed power density above threshold. The white bar in (a)--(c) is $\SI{10}{\mu m}$ and the circular pump radius is $R_{0}=\SI{11.5}{\mu m}$.}
\label{fig_particles}
\end{figure}

\begin{figure}[t]
\includegraphics[scale=1.0]{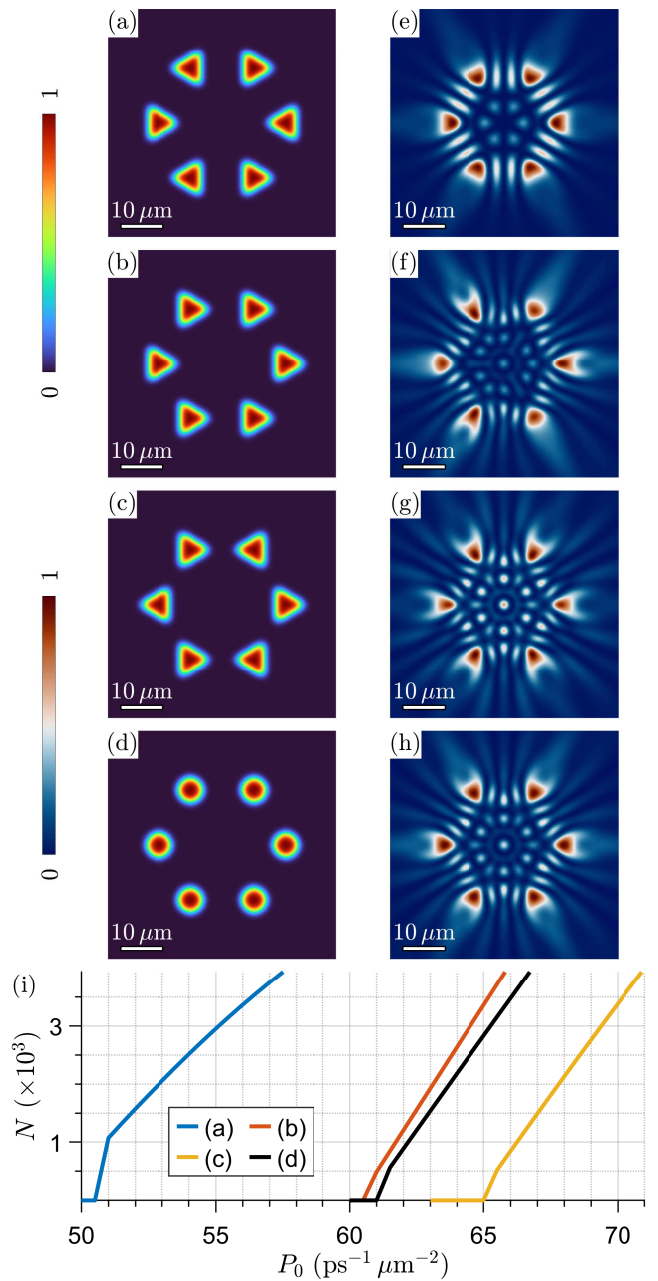}
\phantomsubfloat{\label{fig_layout_a}}
\phantomsubfloat{\label{fig_layout_b}}
\phantomsubfloat{\label{fig_layout_c}}
\phantomsubfloat{\label{fig_layout_d}}
\phantomsubfloat{\label{fig_layout_e}}
\phantomsubfloat{\label{fig_layout_f}}
\phantomsubfloat{\label{fig_layout_g}}
\phantomsubfloat{\label{fig_layout_h}}
\phantomsubfloat{\label{fig_layout_i}}
\vspace{-2.3\baselineskip}
\caption{Pump profiles structured into a hexagon with different relative orientation of the triangular spots: (a) side-side-, (b) side-vertex-, and (c) vertex-vertex-facing nearest neighbors. (d) A reference hexagon of circular spots. (e)--(h) Corresponding normalized condensate densities at $P_0=1.1P_\text{th}$. (i) Corresponding number of condensate particles for increasing power density marking the different condensation thresholds for each configuration.}
\label{fig_layout}
\end{figure}
After calibrating the system, we quantify the anisotropy of the ballistic flow from the condensate steady state by integrating the particle density $N = \int |\Psi|^2 d\mathbf{r}$ over two segmented regions, denoted $N_A$ and $N_B$, shown schematically in Figs.~\ref{fig_particles_a} and~\ref{fig_particles_b} for varying curvature $\kappa = 1/R$ of the triangle spot sides. For the circular pump, the curvature satisfies $\kappa R_0 = 1$ where $R_0$ is its radius, whereas for the circumscribed triangle, $\kappa = 0$ [see Fig.~\ref{fig_particles_c}]. For a cylindrically symmetric spot, the ratio is $N_A/N_B =1$ [see Fig.~\ref{fig_particles_d}], as expected since the condensate steady state also becomes cylindrically symmetric. Approaching the equilateral triangle shape, $\kappa \to 0$, the ratio increases dramatically to almost $N_A/N_B \approx 16$, underlining the strong focusing of the ballistic polariton outflow from the sides of the triangular pump spots. 
\begin{figure*}[t]
\includegraphics[scale=1.0]{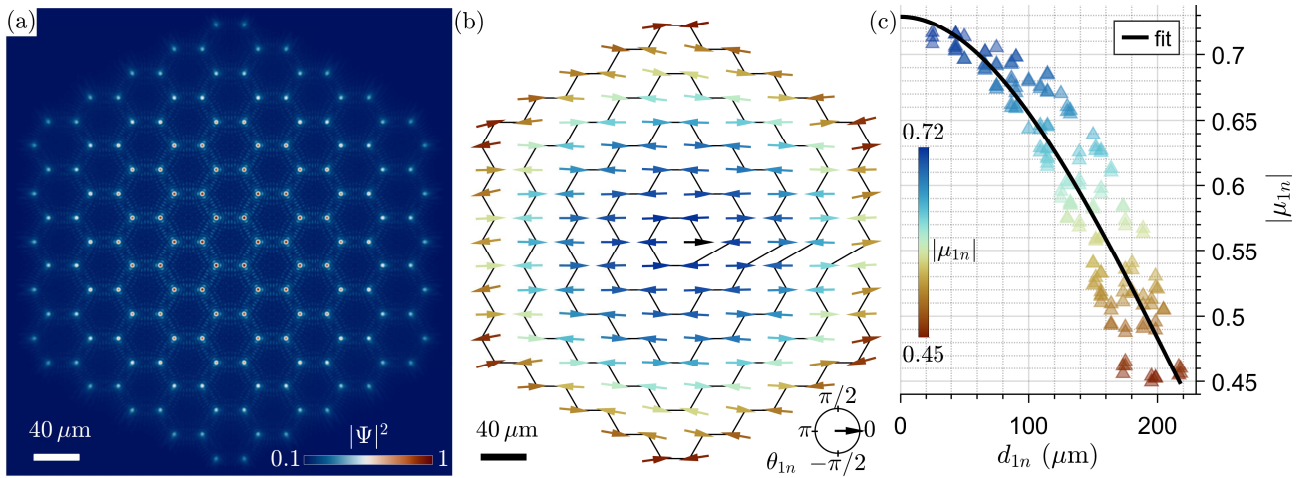}
\phantomsubfloat{\label{fig_single_lattice_a}}
\phantomsubfloat{\label{fig_single_lattice_b}}
\phantomsubfloat{\label{fig_single_lattice_c}}
\vspace{-2\baselineskip}
\caption{(a) Normalized time-integrated condensate density $\langle |\Psi|^2 \rangle$ for triangular pump spots arranged into a honeycomb lattice. (b) Corresponding extracted mutual complex coherence function $\mu_{1n}$. The color scale and arrow orientation depict the magnitude $|\mu_{1n}|$ and phase $\theta_{1n}$, respectively. The central black arrow indicates the reference spot with zero phase. (c) The modulus of the coherence function as a function of absolute distance between the first and $n$-th spots $|\mathbf{r}_1 - \mathbf{r}_n| = d_{1n}$. The black line is a fit of the stretched exponential function~\ref{effective_coherence_length} which gives effective coherence length of $L_{\mathrm{coh}}=\SI{279.3}{\mu m}$.}
\label{fig_single_lattice}
\end{figure*}

Next, we characterize the coupling strength between adjacent condensates pumped with triangular spots of different relative orientation. We focus on four distinct hexagonal pumping patterns [Figs.~\ref{fig_layout_a}--\ref{fig_layout_d}] and their corresponding condensate densities $|\Psi|^2$ at $P_0=1.1P_\text{th}$ [Figs.~\ref{fig_layout_e}--\ref{fig_layout_h}]. The first three patterns can be categorized as side-side-, side-vertex-, and vertex-vertex-facing triangles. In {\color{blue}Fig.~\ref{fig_layout_i}}, we show the population of the condensate  $N = \int |\Psi|^2 \, d\mathbf{r}$ when scanning the pumping power density in time (linearly) while numerically integrating Eq.~\eqref{gross-pitaevskii_equation}. The results show that the lowest threshold belongs to the side-side-facing pattern in Fig.~\ref{fig_layout_a}, whereas the highest threshold belongs to the vertex-vertex-facing pattern in Fig.~\ref{fig_layout_c}. This result intuitively makes sense because the polariton outflow is strongest from the sides of the triangular pump spots which enhances the overlap and coupling between neighbors, and weakest from the vertices, in agreement with the results from Figs.~\ref{fig_isotropy} and~\ref{fig_particles}.

\section{Spatial coherence enhancement}
\label{Spatial coherence in a honeycomb lattice}
The enhanced coupling between side-side-facing triangular pump spots shown in Fig.~\ref{fig_layout} implies stronger spatial coherence in an extended system of polariton condensates which is an essential property to study large-scale emergent phenomena such as macroscopic vorticity~\cite{Cookson_NatComm2021}, universal behaviors and Kibble-Zurek scaling~\cite{caputo2018topological}, and simulation of spin systems~\cite{Berloff_NatMat2017, Tao_NatMat2022}. Here, we demonstrate this enhancement of the condensate coherence length by tiling a large honeycomb lattice of side-side-facing triangular pump spots like in Fig.~\ref{fig_layout_a}. The resulting condensate solution is shown in Fig.~\ref{fig_single_lattice_a}.

In order to calculate the mutual coherence between any two spatial locations of the condensate in the lattice, we use a stochastic generalized Gross-Pitaevskii equation in the truncated Wigner approximation~\cite{Wouters_PRB2009}. This leads to a simple Langevin-type equation for the condensate dynamics which is sufficient to extract the relative improvement between triangular and circular pumping configurations. We note that the truncated Wigner approximation remains valid as long as $\gamma\gg g/\Delta A$, where $\Delta A$ is the simulation grid pixel area. A complex white-noise operator $i \hbar dW/dt$ is appended to Eq.~\eqref{gross-pitaevskii_equation}, representing small random fluctuations added at every time step, with correlators satisfying
\begin{align}
\langle dW_{i}dW_{j} \rangle & = 0,
\\
\langle dW^{*}_{i}dW_{j} \rangle & = \frac{\gamma+Rn_X}{2\Delta A}dt\delta_{i,j}.
\label{correlation_function_white_noise}
\end{align}
Here, $i$ and $j$ refer to different spatial grid points in the numerical simulation and $\Delta A = \Delta x \Delta y$ is the area of the grid cells.

The spatial coherence across the lattice is quantified using the normalized complex first-order coherence function (sometimes denoted as $g^{(1)}$) between each pair of condensates written as
\begin{eqnarray}
\mu_{nm}=\frac{\langle \psi_{n}^{*}\psi_{m} \rangle}{\sqrt{\langle \psi_{m}^{*}\psi_{m} \rangle \langle \psi_{n}^{*}\psi_{n} \rangle}}, \quad n,m = 1,2,\dots 
\label{complex_coherence_factor}
\end{eqnarray}
where $\psi_n(t) = \Psi(\mathbf{r}_n,t)$ is the phase and amplitude of the $n$th condensate at the center of their respective pump spot location $\mathbf{r}_n$. The time average is defined as $\langle \psi_{m}^{*}\psi_{n} \rangle = \frac{1}{T}\int_{T}\psi_{m}^{\ast}\psi_{n} \, dt$, where $T$ is the duration of the simulation which is taken to be much greater than any other characteristic timescale in the model parameters. It is worth noting that in experiment, the mutual coherence function can be measured through multislit interferometry~\cite{topfer2021engineering}. The modulus of the first-order coherence function $|\mu_{nm}| \leqslant 1$ serves as a normalized measure of coherence between any two condensates in the lattice, whereas its argument represents their average phase difference,
\begin{equation}
\theta_{nm}=\mathrm{arg}(\mu_{nm}).
\end{equation}
\begin{figure*}[t]
\includegraphics[scale=1.0]{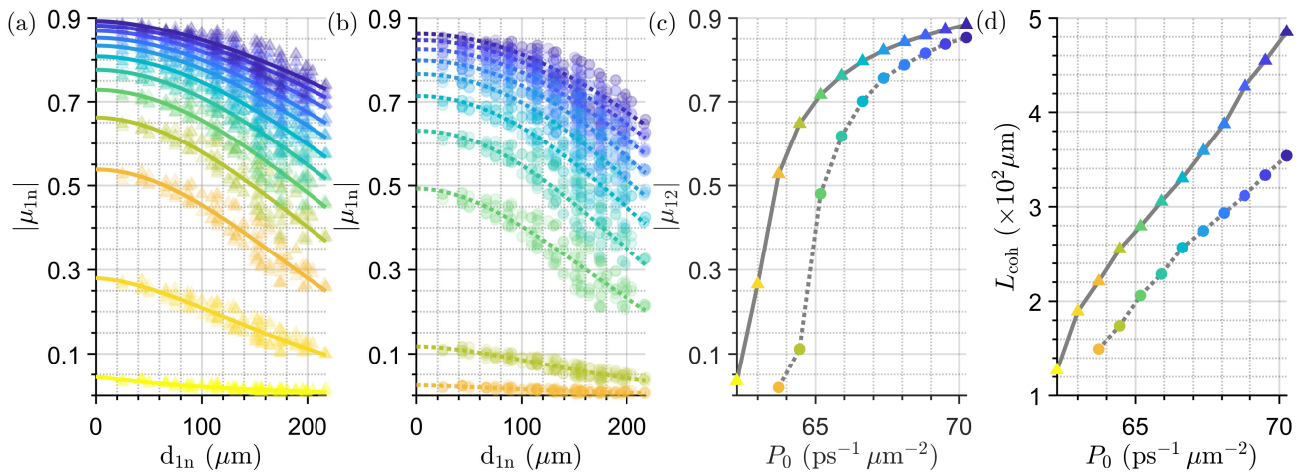}
\phantomsubfloat{\label{fig_multi_lattice_a}}
\phantomsubfloat{\label{fig_multi_lattice_b}}
\phantomsubfloat{\label{fig_multi_lattice_c}}
\phantomsubfloat{\label{fig_multi_lattice_d}}
\vspace{-2\baselineskip}
\caption{The modulus of the mutual coherence function $|\mu_{1n}|$ for increasing absolute condensate neighbor distance $d_{1n}$ and different power densities (colors) for (a) triangular spots and (b) circular spots. Solid and dotted lines are fit using a stretched exponential function. (c) Corresponding power density scan of the mutual coherence between the first and the second condensate $|\mu_{12}|$ for both configurations. (d) Corresponding effective coherence length extracted from fitting for both configurations. The color of the markers directly corresponds to the values on the horizontal axis in (c).}
\label{fig_multi_lattice}
\end{figure*}
%
%
Figure~\ref{fig_single_lattice_a} shows an example condensate time-integrated density for a finite-size honeycomb lattice of side-side-facing triangular spots at a given lattice constant and pump power. Clear interference fringes can be seen between the pumped condensate bright spots, implying robust synchronization even in the presence of noise. Closer to the edge of the lattice, transverse losses due to the strong polariton outflow are more effective, which results in weakened edge density.

In Fig.~\ref{fig_single_lattice_b}, the arrows denote the phase $\theta_{1n}$ between all pairs of condensates with respect to the central one (denoted with a black arrow) with a color scale depicting the coherence amplitude $|\mu_{1n}|$. As expected, the coherence drops radially because of the decreased coupling between distant neighbors. Note that the arrows have arranged themselves antiparallel with respect to nearest neighbors, which implies antiphase ($\pi$) synchronization between the condensates for the given lattice parameters. Other lattice parameters can result in a condensate solution characterized by in-phase synchronization between the lattice nodes~\cite{Topfer_CommPhys2020}. In either case, our conclusions remain the same. We also point out the slight twist in the arrow angles at the edge of the lattice shown in Fig.~\ref{fig_single_lattice_b}. This twist was observed recently in experiment~\cite{topfer2021engineering} and stems from the polariton flowing out of the lattice, which corresponds to a phase gradient between the condensates.

In Fig.~\ref{fig_single_lattice_c}, we plot the modulus of the coherence function $|\mu_{1n}|$ for increasing absolute distance between the central condensate node and the rest, $|\mathbf{r}_1 - \mathbf{r}_n| = d_{1n}$. We obtain a good fit using a stretched exponential function~\cite{caputo2018topological} (black curve), written as
\begin{eqnarray}
\mu(d)=Ae^{-\left(d/B\right)^{C}}, \quad d\geqslant 0.
\label{stretched_exponential_function}
\end{eqnarray}
Here, $A,B,C$ are fitting parameters. Integrating $\mu(d)/A$ from $0 \to \infty$, we obtain an expression for the effective coherence length of the system~\cite{caputo2018topological},
\begin{equation}
L_{\mathrm{coh}}=\frac{B}{C} \times \Gamma\left(C^{-1}\right),
\label{effective_coherence_length}
\end{equation}
where $\Gamma$ is the gamma function. Equation~\eqref{effective_coherence_length} can be regarded as the spatial relaxation length of first-order correlations in the condensate.


We repeat the calculation for the lattice of triangles from Fig.~\ref{fig_single_lattice_c}, but now for several different power densities collated into Fig.~\ref{fig_multi_lattice_a} in different colors. Yellow is the weakest power and blue is the strongest power. For comparison, modulus of the coherence function for a lattice of circular spots is shown in Fig.~\ref{fig_multi_lattice_b}. The comparison is more clearly visualized in Figs.~\ref{fig_multi_lattice_c} and~\ref{fig_multi_lattice_d}, where we plot only the coherence between the central nearest-neighbor condensates $|\mu_{12}|$ and the effective coherence length $L_\text{coh}$, respectively. 
The former [Fig.~\ref{fig_multi_lattice_c}] shows that the condensate coherence is stronger for triangular pump spots across all powers, as expected. It also displays a sharp increase in both cases followed by a saturation, which is similar to past observations~\cite{Deng_PRL2007,topfer2021engineering}. At even higher powers (not shown here), the coherence starts dropping as the condensate becomes unstable and starts fragmenting into multiple energy components~\cite{Topfer_CommPhys2020}. This result underlines the enhanced spatial coupling between triangularly pumped condensate nodes as compared to circularly pumped nodes. Note that the amplitude of the noise for a given power density $P_0$ is the same for triangular and circular spots according to Eq.~\eqref{correlation_function_white_noise} since $n_X \propto P_0$. Therefore, the increased coherence for the triangular spots cannot be attributed to different levels of noise, as compared to circular spots, but rather the focused ballistic emission of polaritons between nearest-neighbor condensate nodes in the lattice. 

The latter [Fig.~\ref{fig_multi_lattice_d}] verifies that not only has the condensate coherence increased using triangular pump spots, but the relaxation of spatial correlations in the condensate lattice is much slower, implying longer coherence lengths, than for the circular spots across all powers tested. On average, over all pump powers tested here, the improvement in coherence is $\approx36\%$. A more exhaustive numerical study over the parameter space of the model will help to accurately quantify the improvement.

\section{Conclusions}
We have shown that by tailoring the shape of an incident nonresonant light source which excites ballistic exciton-polariton condensates, we can enhance the spatial coupling between separately pumped condensates. The reduced symmetry of the tailored pump spots, as compared to typical Gaussian spots, refracts and focuses outflowing high-momentum polaritons from their pumped condensate centers. The coherent flow of polaritons can be focused towards nearest neighbors to enhance condensate spatial coupling. We verify our method by numerically solving the stochastic generalized Gross-Pitaevskii equation for a honeycomb lattice of triangular pump spots which displays a lowered threshold and larger effective coherence lengths as compared to a lattice of circular (cylindrically symmetric) pump spots. Our method can be applied on today's optical microcavities using standard spatial light modulator technology to generate macroscopic fluids of light with improved coherence scales. It can be used to help with exploration into complex long-range dynamics in dissipative quantum fluids, design of large-scale structured coherent light sources in the strong-coupling regime, and development of less noisy analog computing platforms based on polariton networks~\cite{Kavokin_NatRevPhys2022}.
\\

All data supporting this article are available on the University of Southampton's online repository~\cite{datasets_reference}.

\begin{acknowledgments}
The authors acknowledge the support of the European Union’s Horizon 2020 program, through a FET Open research and innovation action under the Grant Agreements No. 899141 (PoLLoC) and No. 964770 (TopoLight). H.S. acknowledges the Icelandic Research Fund (Rannis), Grant No.~217631-051. Y.W.’s studentship was financed by the Royal Society, Grant No. RGF$\backslash$EA$\backslash$180062.
\end{acknowledgments}

\end{document}